\documentclass[conference]{IEEEtran}
\usepackage[utf8]{inputenc}
\usepackage{amsfonts,amssymb,amsmath}
\usepackage{cite}
\usepackage{amsmath,amssymb,amsfonts}
\usepackage{algorithmic,algorithm}
\usepackage{graphicx}
\usepackage{textcomp}
\usepackage{verbatim}
\usepackage{ulem}
\usepackage{booktabs}

\def\BibTeX{{\rm B\kern-.05em{\sc i\kern-.025em b}\kern-.08em
    T\kern-.1667em\lower.7ex\hbox{E}\kern-.125emX}}
\usepackage[usenames]{xcolor}

\makeatletter
    \newcommand{\linebreakand}{%
      \end{@IEEEauthorhalign}
      \hfill\mbox{}\par
      \mbox{}\hfill\begin{@IEEEauthorhalign}
    }
\makeatother

\def\RR{{\mathbb R}}

\def\Id{\mathsf{I}}

\def\be{\begin{equation}}
\def\beq#1{\begin{equation}\label{#1}}
\def\ee{\end{equation}}
\def\bea{\begin{eqnarray}}
\def\beqa#1{\begin{eqnarray}\label{#1}}
\def\eea{\end{eqnarray}}
\def\bean{\begin{eqnarray*}}
\def\eean{\end{eqnarray*}}
\def\ba{\begin{array}}
\def\ea{\end{array}}

\def\argmin{\mathop{\mathsf{arg\, min}}}

\DeclareMathAlphabet{\mathpzc}{OT1}{pzc}{m}{it}

\def\cE{{\mathcal E}}

\def\cN{{\mathcal N}}

\def\cV{{\mathcal V}}

\def\R{\RR}

\def\Tr#1{\mathsf{Tr}\left(#1\right)}

\def\bPsi{\boldsymbol{\Psi}}

\def\ba{{\mathbf a}}

\def\bB{{\mathbf B}}

\def\bG{{\mathbf G}}
\def\bGamma{\mathbf{\Gamma}}

\def\bS{{\mathbf S}}

\def\bW{{\mathbf W}}

\def\bX{{\mathbf X}}

\def\bZ{{\mathbf Z}}

\title{Sparse wavelet-based solutions for the M/EEG inverse problem\thanks{BMWs}}

\author{
\IEEEauthorblockN{Samy Mokhtari\IEEEauthorrefmark{1}, Jean-Michel Badier\IEEEauthorrefmark{2}, Christian B\'enar\IEEEauthorrefmark{2}, Bruno Torr\'esani\IEEEauthorrefmark{1}}
\IEEEauthorblockA{\IEEEauthorrefmark{1} \textit{Aix Marseille Univ, CNRS, I2M, Marseille, France.} \\
\IEEEauthorrefmark{2} \textit{Aix Marseille Univ, CNRS, INS, Inst Neurosci Syst, Marseille, France.} \\
\{samy.mokhtari, jean-michel.badier, christian.benar, bruno.torresani\}@univ-amu.fr}
}

\date{\today}

\begin{document}

\maketitle

\begin{abstract}
This paper is concerned with variational and Bayesian approaches to neuro-electromagnetic inverse problems (EEG and MEG). The strong indeterminacy of these problems is tackled by introducing sparsity inducing regularization/priors in a transformed domain, namely a spatial wavelet domain. Sparsity in the wavelet domain allows to reach "data compression" in the cortical sources domain. Spatial wavelets defined on the mesh graph of the triangulated cortical surface are used, in combination with sparse regression techniques, namely LASSO regression or sparse Bayesian learning, to provide localized and compressed estimates for brain activity from sensor data. Numerical results on simulated and real MEG data are provided, which outline the performances of the proposed approach in terms of localization.
\end{abstract}

\begin{IEEEkeywords}
MEG inverse problem, compression, spatial wavelets, non-smooth optimization, Sparse Bayesian Learning
\end{IEEEkeywords}

\section{Introduction}
\label{se:intro}
MEG and EEG inverse problems~\cite{Baillet2017magnetoencephalography} are extremely underdetermined, the number of measurements being much smaller than the number of unknowns (by a factor $\sim 50$). In such a case, the obtained solution dramatically depends on the choice of regularization and/or prior information. Reference solutions, routinely used by practitioners, often suffer from known biases. For example, the Minimum-Norm Estimate (MNE) framework tends to overestimate and oversmooth the support of the cortical activity, while other sparsity-enforcing approaches (e.g. Minimum-Current Estimate) can yield very localized solutions, which may be too localized in many situations. Last but not least, the absence of ground truth makes objective comparison of different methods very difficult.

The goal of this paper is to investigate solutions exploiting sparsity in a spatial wavelet domain, with the objective of describing spatially extended (i.e. patch-like) solutions without prior assumption on the patch size. Limiting to a particular instance of spatial wavelet transform, i.e. spectral graph wavelets defined on the triangulated cortical surface, we explore a wavelet-based MNE solver paired with a Sparse Bayesian Learning approach, and analyze its performances in comparison with some state of the art methods : MNE, MCE, and sparse total variation penalization.

\section{Problem statement and state of the art}
\label{se:SOTA}
Given a MEG (magnetoencephalography) dataset consisting in $J_{0}$ sensors measurements during $L$ time steps, and summarized in the form of a matrix $\mathbf{Z} \in \R^{J_{0} \times L}$, MEG inverse problem consists in estimating the underlying cortical current densities $\mathbf{S} \in \R^{N\times L}$ explaining (in an "optimal" way) the data. Typical orders of magnitude for the dimensions at stake are $J_{0} \approx 250$ and $N \approx 15,000$. With a suitable whitening matrix $\mathbf{\Upsilon} \in \R^{J_{0} \times J}$ and dimension reduction onto $J < J_{0}$ spatial channels, the measurements and cortical sources are connected by the following observation equation 
%
%
\begin{equation}
\label{fo:observation.equation}
\mathbf{Z} = \mathbf{G} \mathbf{S} + \mathbf{B}
\end{equation}
with $\mathbf{Z} = \mathbf{\Upsilon} \mathbf{Z_{0}}$ the whitened data, $\mathbf{G} = \mathbf{\Upsilon} \mathbf{G_{0}}$ the whitened leadfield matrix, and $\mathbf{B}$ the matrix of observation noise and background activity, assumed i.i.d. with zero mean and unit variance (as a result of whitening). The leadfield matrix $\mathbf{G}$ summarizes the propagation of the electromagnetic field from cortex to sensors. 

We here focus on variational and Bayesian approaches for distributed source models with constrained orientation (see~\cite{Baillet2001electromagnetic}). Reference approaches also include dipole fitting~\cite{Mosher1992multiple,Jerbi2004localization} and various forms of beamformers~\cite{VanVeen1997localization,Sekihara2008adaptive}. Variational methods generally lead to optimization problems involving a quadratic data fidelity term and a penalty term $f( \mathbf{S} )$ :
\begin{equation}
\label{fo:inverse.problem}
\mathbf{S_*} = \argmin_{ \mathbf{S} \in \RR^{N\times L}} \left[\frac{1}{2}\left\| \mathbf{Z} - \mathbf{G} \mathbf{S} \right\|_F^2 + f( \mathbf{S} )\right]\ ,
\end{equation}
where $\|\cdot\|_F$ stands for the Frobenius norm.

Quadratic regularizations $f( \mathbf{S} )=\| \mathbf{D} \mathbf{S} \|_F^2$ yield closed-form solutions ($\mathbf{D}=\sqrt{\lambda}\, \mathbf{\Id}$ for MNE~\cite{Hamalainen1994interpreting} and $\mathbf{D}$ diagonal for weighted MNE~\cite{Fuchs1999linear}). These methods turn out to be strongly affected by \textit{source leakage} (oversmoothing and overestimation of spatial extent) and \textit{depth bias} (under-estimation of deep sources).

The strong under-determination of the optimization problem (\ref{fo:inverse.problem}) has pushed for the introduction of sparsity inducing regularization/priors, which are implemented in variational formulations through a non-smooth, generally convex regularization $f( \mathbf{S} )$. Successful examples include the minimum current estimate (MCE ~\cite{Matsuura1995selective}, $f( \mathbf{S} )=\lambda\| \mathbf{S} \|_1$), mixed norm estimate that promote sparsity in space and persistence in time (MxNE~\cite{Ou2009distributed,Gramfort12MixedNorm}, $f( \mathbf{S} )=\lambda\| \mathbf{S} \|_{21}$), total variation (VB-SCCD~\cite{Ding2009reconstructing}, $f( \mathbf{S} )=\lambda \|\nabla \mathbf{S} \|_1$), sparse total variation (sVB-SCCD~\cite{Becker2014Fast}, $f(\mathbf{S})=\lambda \|\nabla \mathbf{S} \|_1 + \mu \|\mathbf{S}\|_1$) and similar approaches involving time persistence. The problem is then solved through dedicated numerical algorithms (see~\cite{Becker2014Fast} for a review).

These methods involve regularization parameters, whose tuning and interpretation may rely on a Maximum A Posteriori (MAP) estimation in a Bayesian point of view. Quite recently, empirical Bayes methods such as Sparse Bayesian Learning (SBL, see~\cite{Wipf2010robust} and~\cite{Hashemi2021unification} for a review) have received increasing interest in the M/EEG inverse problem literature for their ability to enforce sparse solutions. SBL involves joint optimization of the objective function in~\eqref{fo:inverse.problem} in the case $f(\mathbf{S})=\|\mathbf{\Gamma}_{S}^{-1/2} \mathbf{S} \|_F^2$, with $\bGamma_\bS=\mathsf{diag}(\underline{\gamma_{S}})$ and $\underline{\gamma_{S}}\in\RR^N$, and treats $\bS$ as a latent variable. Marginalizing with respect to $\bS$ results in the non-convex optimization problem
\begin{equation}
\label{fo:SBL_non_convex_optimization_problem}
\underline{\gamma_{S}}^{*}\!=\!\argmin_{\underline{\gamma_{S}} \in \left( \R_+ \right) ^N}\! \left[\Tr{ \mathbf{C_Z} \mathbf{\Sigma_Z} (\underline{\gamma_{S}})^{-1}} \!+\! \ln\det( \mathbf{\Sigma_Z} (\underline{\gamma_{S}}))\right]\, ,
\end{equation}
where $\mathbf{\Sigma_Z} ( \underline{\gamma_{S}} ) = \mathbf{\Id_J} + \mathbf{G} \mathbf{\Gamma}_{S} \mathbf{G}^T$ is the posterior data covariance matrix, and $\mathbf{C_Z} = \frac{1}{L}\mathbf{Z} \mathbf{Z}^T$ the sample covariance matrix. Given $\underline{\gamma_{S}}^{*}$, the corresponding  weighted MNE solution reads :
\begin{equation}
    \mathbf{S}_{*} = \mathbf{\Gamma}_{S}^{*} \mathbf{G}^{T} \left[ \mathbf{\Sigma}_{Z}(\underline{\gamma_{S}}^{*}) \right]^{-1} \mathbf{Z}
\end{equation}
The so-obtained solutions can be shown to be sparse, with support size not exceeding the rank of $\bG$.

\section{Spatial wavelet based approaches}
\label{se:WavApp}

Sparse solutions such as MCE or SBL have been proven efficient for estimating very focal brain activity, but tend to underestimate the spatial support in the case of more extended activity. Enforcing sparsity in a transformed domain has been shown to improve the situation (see e.g.~\cite{Becker2014Fast}, where spatial gradient is used as transformation). We focus here on spatial wavelet transforms, which provide multiscale representations for functions on the cortical surface or graph. Several constructions of such wavelets have been proposed in literature. In the current work, we focus on the Spectral Graph Wavelets (SGW) construction of~\cite{Hammond2011wavelets} which, unlike several other constructions, provides a redundant multiscale representation.

\subsection{Laplacian and gradient on the cortical surface, discrete total variation and spectral graph wavelets}

Available data for the M/EEG inverse problem generally involve a discretization of the cortical surface, in the form of a triangular mesh, which may be described as a weighted graph $(\cE,\cV,w)$, with edge set $\cE=\{e_1,\dots e_E\}$, vertex set $\cV=\{v_1,\dots v_N\}$, and a weight function $w\text{ : }\cE \longmapsto \R_{+}$. This defines a symmetric adjacency matrix $\mathbf{A} \in \R^{N \times N}$, whose nonzero elements are associated with neighboring vertices, and are defined by $\mathbf{A}_{kk'}=w(e_{k,k'})$, $e_{k,k'}$ being the edge connecting vertices $k$ and $k'$.
A gradient matrix $\mathbf{\nabla} \in\RR^{E\times N}$ may also be defined on the graph as follows: for any edge $e=e_{k,k'}$, set $\mathbf{\nabla}_{ek} = \mathbf{A}_{kk'}$  and $\mathbf{\nabla}_{ek'}=-\mathbf{A}_{kk'}$. The corresponding TV norm of $\mathbf{S} \in \RR^N$ is then defined as $\|\nabla \mathbf{S}\|_1$. This notion is used in the VB-SCCD and sVB-SCCD solvers.

The un-normalized graph Laplacian is then defined as the $N \times N$ matrix $\mathbf{L}=\mathbf{D}-\mathbf{A}$, where $\mathbf{D}=\mathsf{diag}\{d_1,\dots d_N\}$ is the degree matrix: $d_{k}= \sum_{k'} \mathbf{A}_{kk'}$, the sum running over neighboring vertices. $\mathbf{L}$ is symmetric positive semi-definite. Its eigenvectors $\{\chi_n\}_{n=1}^{N}$ may thus be interpreted as Fourier modes for functions defined on the graph, and the corresponding eigenvalues $0=\lambda_1\le\dots\le\lambda_N$ as (spatial) frequencies. Denoting by $f\mapsto \widehat{f}$ the corresponding Graph Fourier Transform, a scaling operator $T_{h}$ and a wavelet operator $T_{g}^s$ on $\RR^N$ can be defined~\cite{Hammond2011wavelets} as follows: for all $f \in \R^{N}$ and all Fourier modes $l \in \{1,\hdots,N\}$
%
\begin{equation}
\widehat{T_{h} [f]} (l) = h ( \lambda_{l} ) \left \langle \chi_{l}, f \right \rangle, \quad
\widehat{T_{g}^{s} [f]} (l)  =  g ( s \lambda_{l} ) \left \langle \chi_{l}, f \right \rangle \ ,
\end{equation}
%
where $h$ and $g$ denote respectively a low-pass and band-pass kernel functions defined on the continuous extension $\RR_+$ of the spectrum of $\mathbf{L}$, and $s \in \R_{+}^{*}$ refers to a dilation parameter.
Scaling functions $\phi_n$ and scaled wavelets $\psi_{s,n}$ can be defined as $\phi_n=T_h[\delta_n]$ and $\psi_{s,n}=T_g^s[\delta_n]$ respectively, $\delta_n$ being the Dirac mass at vertex $n$. For suitably chosen $g,h$ and scales $\{s_1,\dots s_{N_s}\}$, these functions form a frame of $\RR^N$, of cardinality $N_W=N(N_s+1)$. The corresponding matrix reads
%
%
\begin{equation}
\mathbf{W}  = \begin{pmatrix}
\phi_{1} & \hdots & \phi_{N} & \psi_{s_{1},1} & \hdots & \psi_{s_{N_{s}},N} \\
\end{pmatrix}^{T} \ .
\end{equation}
This matrix $\mathbf{W}$ allows defining the wavelet analysis operator $\mathcal{W}$, whose adjoint $\mathcal{W}^{*}$ is the synthesis operator with matrix $\mathbf{W}^{T}$. 
In synthesis based approaches, a function $f$ on the graph is expressed as $f=\bW^T\bX$.

\medskip

It is here worth highlighting that the cutoff frequency of the low-pass kernel $h$, the number of detail coefficients $N_{s}$, and the wavelets quality factor are hyper-parameters of the wavelet-based method hereafter described.



\subsection{Synthesis-based variational formulation}

Given this spectral graph wavelet construction, we introduce sgw-MNE, the analogue of the MNE problem~\eqref{fo:inverse.problem}. From a synthesis point of view, writing $\bS=\bW^T\bX$ and $\mathbf{G_{W}} = \mathbf{G} \mathbf{W}^{T}$:
\begin{equation}
\label{fo:wavelet_MNE_inverse_problem}
\mathbf{X}_* = \argmin_{ \mathbf{X} \in \R^{N_W \times L}} \left[\frac{1}{2}\left\| \mathbf{Z} - \mathbf{G_{W}} \mathbf{X} \right\|_F^2 + \lambda \| \mathbf{X} \|_F^2 \right]\ ,
\end{equation}
%
The problem has a closed-form solution $\bX_*$, and the corresponding source estimate then reads $\bS_*=\bW^T\bX_*$.

As for the above hyper-parameter $\lambda$, it can be estimated, thanks to the Bayesian point of view, from the signal to noise ratio (SNR). Indeed, defining the signal and noise levels in the sensor domains by the traces of the corresponding covariance matrices $\mathbf{\Sigma_{Z}} = \mathbf{\Sigma_{B}} + \mathbf{G} \mathbf{\Gamma_{S}} \mathbf{G}^T$ and $\mathbf{\Sigma_{B}}$, the corresponding SNR $\rho$ (ratio of signal and noise standard deviations) satisfies $\rho^2 = 1+\|\bG\|_F^2/\lambda\Tr{\mathbf{\Sigma_{B}}}$.
%
%
For the sake of completeness, a sgw-MCE solution can also be obtained by replacing the quadratic penalization in~\eqref{fo:wavelet_MNE_inverse_problem} with a term $\lambda\|\bX\|_1$, and a similar heuristics can be derived to connect $\lambda$ to $\rho$ in the MCE case.

\subsection{Sparse Bayesian Learning formulation}

To make up for the excessive sparsity enforced by a straightforward SBL approach, we designed a SBL algorithm that enforces sparsity in the graph wavelet domain instead of the spatial domain. 
The implementation of the SBL approach within the wavelet-based variational formulation (\ref{fo:wavelet_MNE_inverse_problem}) is straightforward, as it follows an optimization problem similar to (\ref{fo:SBL_non_convex_optimization_problem}), with now $\mathbf{\Sigma_Z} ( \underline{\gamma_{X}} ) = \mathbf{\Id_J} + \mathbf{G_W} \mathbf{\Gamma}_{X} \mathbf{G_W}^T$.\newline
\indent Several algorithms are described in~\cite{Hashemi2021unification} for minimizing such objective functions, such as Expectation-Maximization and convex-bounding Champagne and variants. They all require an initial value for $\underline{\gamma_{X}}$, which we obtain from a preliminary sgw-MNE computation followed by the evaluation of the estimated source variances. We term this approach sgw-SBL.

\section{Results}
\label{se:NumRes}

The proposed approach is first illustrated on simulated data, numerically generated on a brain geometry based on a MEG dataset acquired by the DYNAMAP group at INS Marseille in an auditory evoked potentials protocol. Available data include sensor data (245 MEG sensors, 180 trials, sampled at 2034.51 Hz). 
%
%
The dataset also involves a triangulated cortical surface (interface of white and gray matters), with 14995 vertices. 3D coordinates are available, as well as the adjacency matrix of the mesh graph, which permits  the computation of graph Gradient and Laplacian. A leadfield matrix $\mathbf{G}$ was computed using the OpenMEEG software.
A baseline covariance matrix $\mathbf{\Sigma}_{B}$ was estimated from pre-stimulus data, 
and used for data pre-whitening, and projection onto the subspace spanned by eigenvectors with significant eigenvalues.\newline
\indent The simulation protocol is as follows : $N_{p} = 100$ connected patches on the cortical surface were generated with random locations across the brain, with two different sizes (10 and 100 vertices per patch). For each patch, simulated sources $\bS_\mathsf{sim}$ were set to a constant value $\beta$ inside the patch and zero outside. Corresponding sensor data were computed as $\bZ_\mathsf{sim}\!=\!\bG\bS_\mathsf{sim}\!+\! B$, with $B$ a realization of sensor baseline data, simulated as iid samples from a multivariate normal distribution $\cN(0,\mathbf{\Sigma}_\bB)$. The constant $\beta$ was tuned to match a given \textsf{\small PSNR} (ratio of peak signal value to noise standard deviation), in the sensor domain. After whitening and dimension reduction, source estimates were computed with several approaches : MNE, sgw-SBL, MCE and sVB-SCCD.

\subsection{Wavelet design}

In this paper, we use a mesh graph with binary weights, and limit ourselves to the SGW design described in section 8.1 of~\cite{Hammond2011wavelets}, with a bump shaped, piecewise polynomial, band-pass kernel $g$, and a low-pass kernel $h$ with very fast decay. The cutoff frequency of the low-pass kernel is set to $\lambda_{min}=\lambda_{max}/K$, with $\lambda_{max}$ the largest eigenvalue of the Laplacian $\mathbf{L}$, and $K>1$ a user-defined constant. The quality factor (center frequency divided by bandwidth) of the bandpass kernel approximately equals $1.38$. The wavelets frame bounds are approximately equal to $0.71$ and $1.41$.

\subsection{Evaluation metrics}

Objective evaluation of MEG source reconstruction methods is a difficult problem. For real data, the absence of ground truth imposes relying on expert knowledge. No real consensus exists either for result evaluation on simulated data. Indeed, many metrics have been introduced in literature, most of which having a tendency to favor specific approaches or cases.

If one is mostly interested in finding the (approximate) spatial support of the brain activity, the obtained solution has to be thresholded, and the result depends on the thresholding strategy. Methods can be assessed using standard statistical tools (ROC or PR curves, AUC,...), which are not necessarily appropriate in this context, or custom metrics. Since we are primarily interested here in quantitative source estimates, and not only support, we will not follow this avenue here.

In the M/EEG inverse problem literature (see for instance ~\cite{Molins2008quantification,Hauk2022towards}), dedicated metrics such as the Dipole Localization Error (DLE), the Spatial Dispersion (SD) and the Resolution Index (RI) are often put forward. These are generally defined for linear source estimation methods where a closed-form solution exists. In~\cite{Samuelsson2021spatial}, slightly different definitions were introduced for the DLE and SD metrics in order to deal with non-linear source estimation methods, but there does not seem to exist a real consensus in literature. In addition, it has also been stressed out that such metrics, while well-suited for evaluating the localization of a single or a few separable sources, become less relevant for extended sources and complex patterns, especially with non-linear methods. \newline
\indent In this paper, due to limited space, we decided to restrict ourselves to the following metrics: 

\begin{itemize}
\item
a Spatial Dispersion at "peak-time" defined as follows :
%
\begin{equation}
\label{fo:spatial_dispersion_definition}
    SD_{i_{max}} \left( t_{max} \right) = \sqrt{\frac{ \sum_{k=1}^{N} d_{i_{max},k}^{2} \vert \mathbf{S}_{k,t_{max}} \vert^{2} } { \sum_{k=1}^{N} \vert \mathbf{S}_{k,t_{max}} \vert^{2} }}\ ,
\end{equation}
where $t_{\max}$ denotes the time achieving maximal value in sensor domain, $i_{\max}$ the vertex with maximal value at $t_{\max}$, and $d_{i,k}$ the Euclidean distance between vertices. 

\vspace{1mm}

\item
the first Wasserstein distance~\cite{Peyre2020computational} between suitably normalized "energy maps", which are defined by considering an average over a time index range of interest $[l_0,l_1]$ :
\begin{equation}
\label{fo:energy_maps_equation}
    S_{amp,k} \propto \sqrt{\frac{1}{l_{1}-l_{0}+1} \sum_{l=l_{0}}^{l_{1}} \mathbf{S}_{k,l}^{2}}\ .
\end{equation}
\end{itemize}


In addition, to assess the ability to recover a correct order of magnitude, we also record the $\ell^2$ norms of amplitudes $S_{amp}$ for each solver, and compare these values with the reference.
\subsection{Simulation results}

We collect in this section a sample of results comparing the various methods. We provide 3 tables including summary statistics (mean and median, and standard deviation and inter-quartile distance) of evaluation metrics described above. For these tables, sgw-SBL results were obtained with a convex-bounding Champagne algorithm. \textsc{Table}~\ref{tab:spatial_dispersion_metric} displays the obtained values for the SD metric (normalized by the reference spatial dispersion) for all four methods under consideration, and for the two simulated patch sizes. Similarly, \textsc{Table}~\ref{tab:statistics_wasserstein_metric} gives values for the Wasserstein distance between the simulated and estimated normalized energy maps. Looking at these two first evaluation metrics, it can be seen that for the large support simulation, sgw-SBL outperforms consistently and significantly the other approaches, while sVB-SCCD also yields acceptable results. For smaller support size simulations, MCE yields better results with respect to the Wasserstein distance, but sgw-SBL is not far behind and even produces good results with respect to the SD metric. This is not really surprizing, since MCE promotes sparsity in the spatial domain, and can therefore be expected to perform well for focal sources. To better challenge MCE in this case, the wavelets hyper-parameters (low-pass cutoff frequency, number of detail coefficients) may be tuned to favor highly localized sources.\newline
\indent The results displayed in \textsc{Table}~\ref{tab:ratio_L2_norm_metric} also advocate, in the large support case, for sVB-SCCD and sgw-SBL when it comes to the ability to recover a correct order of magnitude for the $l^{2}$ norm of the sources amplitudes $S_{amp}$. Interestingly, sgw-SBL is able to avoid any over-estimation of the $l^{2}$ norm.

\vspace{-2mm}
\setlength{\tabcolsep}{5pt}
\begin{table}[h!]
    \centering
    \caption{\centering{Summary statistics for the normalized spatial dispersion : \newline small v./ large support}}
    \label{tab:spatial_dispersion_metric}
    
    \vspace{-2mm}
    \begin{tabular}{ccccc}
    \toprule
    {$\frac{SD_{est}}{SD_{ref}}$} & MNE       &  sgw-SBL                      &     MCE           &  sVB-SCCD \\
    \midrule
    mean                &    6.27 / 2.40      & \textbf{2.67} / 0.72  &    2.79 / \textbf{1.12}    &   2.82 / 1.20 \\
    median                &    5.88 / 2.34      &         2.36 / 0.64           & \textbf{2.09} / 0.97 &  2.30 / 0.94 \\
    \midrule
    std                 &    2.22 / 0.43      & \textbf{1.23} / \textbf{0.29}  &    2.22 / 0.73    &   1.98 / 0.82 \\
IQD &2.44 / 0.5 &1.31 / 0.5 & 2.87 / 0.55 & 1.4 / 0.93\\
    \bottomrule
    \end{tabular}
    \vspace{.5mm}

\end{table}

\vspace{-6mm}

\begin{table}[h!]
    \centering
    \caption{Summary statistics for the Wasserstein metric : \newline small v./ large support}
    \label{tab:statistics_wasserstein_metric}

    \vspace{-2mm}
    \begin{tabular}{ccccc}
    \toprule
      $W_1$   & MNE & sgw-SBL & MCE  & sVB-SCCD \\
         \midrule
    mean     & 6.96 / 1.34    &  6.47 / \textbf{0.88 }    &   \textbf{5.54} / 1.87    &   6.65 / 1.11    \\
    median     & 6.04 / 1.05    &  5.55 / \textbf{0.63}   &   \textbf{4.08} / 1.56      & 5.62 / 0.79 \\
    \midrule
    std      & 5.19 / 0.96    &  5.08 / 0.74     &   4.92 / 1.65        &    5.25 / 0.95    \\
    IQD &4.58 / 0.82 &4.67 / 0.64 & 4.19 / 1.25 & 4.82 / 0.78\\
%
    \bottomrule
    \end{tabular}
    \vspace{.5mm}

\end{table}

\vspace{-6mm}

\begin{table}[h!]
    \centering
    \caption{Summary statistics for the ratio of $L^{2}$-norms : \newline small v./ large support}
    \label{tab:ratio_L2_norm_metric}
    
    \vspace{-2mm}
    \begin{tabular}{ccccc}
    \toprule
    {$\frac{\| S_{amp}^{est}  \ \|_{2}}{ \| S_{amp}^{ref} \|_{2}}$} & MNE & sgw-SBL & MCE  & sVB-SCCD \\
    \midrule
        mean    &    0.04 / 0.13   & 0.14 / 0.54            &  \textbf{1.12} / 3.25 &   0.40 / \textbf{1.09} \\
        median    &   0.04 / 0.13   &  0.14 / 0.55            &  \textbf{1.14} / 3.28 &   0.37 / \textbf{1.08} \\
        \midrule
        std     &   0.01 / 0.02   &  0.02 / 0.08            &  0.45 / 0.97 &    0.19 / 0.43 \\
        IQD & 0.01 / 0.02& 0.02 / 0.1 & 0.6 / 1.27 & 0.2 / 0.45\\
    \bottomrule
    \end{tabular}
    \vspace{.5mm}

\end{table}

\begin{figure}[h!]
    \centering
    \includegraphics[width=\columnwidth]{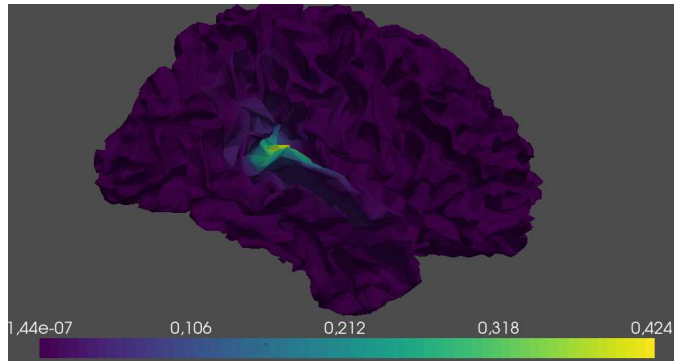}
    \caption{Real data (auditory evoked potentials): estimated sources with sgw-SBL (EM solver)}
    \label{fig:sgw_champagne_real_data}
\end{figure}

To add some experimental basis to this preliminary numerical validation, we also display the solution obtained with a sgw-SBL solver based on an Expectation-Maximization algorithm, with the real data acquired during the auditory evoked potentials protocol. Despite the lack of ground truth solution in this case, Fig. \ref{fig:sgw_champagne_real_data} tends to confirm the performance of sgw-SBL, as the localization of the estimated sources is close to the auditory cortex in the right hemisphere (see~\cite{wikipedia_auditory_cortex}). Furthermore, the order of magnitude of the sources amplitudes is much higher with sgw-SBL compared with what can be obtained with MNE.

\section{Conclusions}
\label{se:Conclu}
We described in this paper sgw-SBL, a spatial wavelet-based method to the MEG inverse problem, where sparsity is enforced in the spectral graph wavelet domain via a sparse Bayesian learning approach. The goal is to estimate spatially extended brain activity, with quantitatively relevant amplitude.

Numerical results on simulated data show that sgw-SBL indeed achieves the best results when the spatial extent of the brain activity is large enough, and even comes as a close second behind MCE for smaller support activity, which are generally better accounted for by approaches favoring more classical spatial sparsity. Such results, well known in image processing, seem to remain true in the non-standard setting considered here (spectral graph wavelets).

Due to space limitation, we restricted the discussion to a limited set of methods. More thorough results investigating a larger family of objective functions (including in particular $l^1$ penalized least square in wavelet domain and analysis-based approaches), a more sophisticated modeling of time dependence, and the influence of hyper-parameters (SGW parameters, regularization parameters...) and SBL algorithms will be presented in a forthcoming publication. 

Further work will also include different wavelet constructions, including SGW with different graph choice (for example taking into account physical edge lengths), or wavelet bases, such as intertwining wavelets on graphs~\cite{avena2018intertwining}, lifting based constructions~\cite{Sweldens1998lifting} or diffusion wavelets~\cite{Coifman2006diffusion} (we recall that SGW is redundant, which induces a significant increase of the problem dimensionality). The integration of spatial wavelets into different Bayesian inference schemes such as Bernoulli-Gauss models~\cite{Barbault2022parameter}, Maximum Entropy on the Mean approach (MEM, see~\cite{Lina2014wavelet,Roubaud2018spacetime}) or Variational Bayes~\cite{Oikonomou2020novel} would also be worth investigating.
\section*{Acknowledgements}
This work has benefited from support from the French government through the BMWs project (ANR-20-CE45-0018), and the \textit{Institut Convergence} ILCB (ANR-16-CONV-0002).

We also wish to thank M. Kowalski, G. Mebarki and C. Mélot for fruitful discussions.

\newpage
\bibliographystyle{myIEEEtran}
\bibliography{MEEG}

\end{document}